\documentclass[fleqn,usenatbib,useAMS]{mnras}

\usepackage{graphicx}	
\usepackage{amsmath}	
\usepackage{amssymb}	
\usepackage{multicol}        
\usepackage{bm}		
\usepackage{pdflscape}	
\usepackage{tabularx}
\usepackage{array}

\usepackage[T1]{fontenc}
\usepackage{ae,aecompl}

\usepackage{newtxtext,newtxmath}

\title[CSS\_J154915.7+375506:a marginal contact binary]{\textit{CSS\_J154915.7+375506: A low-mass-ratio marginal contact binary system with a hierarchical third body}}

\author[Jin-Feng Wu et al.]{
Jin-Feng Wu,$^{123}$\thanks{E-mail: wujinfeng@ynao.ac.cn}
Li-Ying Zhu,$^{123}$\thanks{E-mail: zhuly@ynao.ac.cn}
Azizbek MATEKOV,$^{124}$
Lin-jia Li,$^{13}$
Shuhrat EHGAMBERDIEV,$^{45}$
\newauthor
Ildar ASFANDIYAROV,$^{4}$
Jiang-Jiao Wang,$^{123}$
Jia Zhang$^{13}$
and
Fang-Bin Meng$^{123}$\\
$^{1}$Yunnan Observatories, Chinese Academy of Sciences (CAS), Kunming 650216, China\\
$^{2}$University of Chinese Academy of Sciences, No.1 Yanqihu East Rd, Huairou District, 101408, Beijing, People's Republic of China\\
$^{3}$Key Laboratory of the Structure and Evolution of Celestial Objects, Chinese Academy of Sciences, 650216, Kunming, People's Republic of China\\
$^{4}$Ulugh Beg Astronomical Institute of the Uzbekistan Academy of Science, Astronomicheskaya 33, Tashkent, 100052, Uzbekistan\\
$^{5}$National University of Uzbekistan, 4 University str, Tashkent, 100174, Uzbekistan\\
}

\date{Last updated 2020 June 10; in original form 2013 September 5}

\pubyear{2020}

\begin{document}
\label{firstpage}
\pagerange{\pageref{firstpage}--\pageref{lastpage}}
\maketitle

\begin{abstract}
We presented the multi-filter light curves of CSS\_J154915.7+375506 inaugurally, which were observed by the 1.5 m AZT-22 telescope at
Maidanak Astronomical Observatory. A low-resolution spectrum obtained by LAMOST reveals it is an A-type close binary. By analyzing the
BVRI total-eclipse light curves, we are able to derive a reliable photometric solution for this system, which indicates that CSS\_J154915.7+375506
is an extremely low-mass-ratio (q=0.138) marginal contact binary system. The location in the HR diagram shows that its secondary component with
a much smaller mass is the more evolved one, indicating the mass ratio reversal occurred. The present secondary component had transferred a significant
amount of mass to the present primary one. By the combination of a total of 20 times of minimum, we investigated its O-C curve. A periodic
oscillation and a possible period decrease have been detected. As the period decreases, the system will evolve towards the contact phase.
This makes CSS\_J154915.7+375506 a valuable case to study the formation scenario of contact binaries through mass reversal. The periodic
oscillation suggested a third body with a minimal mass of $0.91\,M_{\odot}$, which is larger than that of the less massive component in the central binary.
This implies that the secondary body was not replaced by the third body during early stellar interactions, indicating that it is a fossil
system and retains its original dynamical information.

\end{abstract}





\section{Introduction}
Eclipsing binary systems exhibiting EB-type light variations, particularly those of short periods, offer captivating insights into the evolutionary
transformations experienced by close binary systems. Near-contact binaries(NCBs), an important subclass of close binaries which lie in a key evolutionary
stage\citep{2006MNRAS.367..423Z}, were defined by\citep{1994AAS...185.8506S}. Considering the geometric structure of NCBs, it can be divided into 4 subclasses:
SD1 type or SD2 type, semi-detached binaries with star1 or star2 filling its critical Roche lobe; D type, detached binary with components very close to their
critical Roche lobe; and C type, marginal contact binary, both components filling their critical Roche lobe with large temperature differences\citep{2009ASPC..404..189Z}.
The marginal contact system is considered a special type of NCB in testing the Thermal relaxation oscillation (TRO) theory
\citep{lucyURSAEMAJORISSYSTEMS1976,lucyObservationalTestsTheories1979,flanneryCyclicThermalInstability1976}. On the basis of TRO theory, the long-term instability of
thermal equilibrium in contact configurations results in the status of marginal contact, together the semi-detached constitute two phases of periodic thermal relaxation oscillation.
NCBs which lie in these two phases that sustain such a short time compared to the lifetime of the binary, are rare. So, despite decades of comprehensive research, the properties and evolution on marginal
contact binaries still remain blurred. A recently formed contact binary will undergo thermal evolution towards a state of marginal contact,
and if the contact is subsequently disrupted, the system will revert back into contact\citep{1976ApJ...205..208L}. This implies that a marginal contact binary could
represent a pre-contact binary system. The transitional nature of marginal contact binaries is a crucial evolutionary stage that has the potential to enhance our understanding of the evolutionary process of NCBs.
Building on this, K. Stepien and M. Kiraga's evolutionary model\citep{2013AcA....63..239S}suggests that contact binaries may have experienced mass ratio reversal in the past.
Based on the discussion of marginal contact binaries above, we will now introduce our upcoming research, which will delve into a specific marginal contact binary.\newline

The short-period near contact binary CSS\_J154915.7+375506 ($\alpha_{J2000}=15^h49^m15^s.7, \delta_{J2000}=+37^{\odot}55^{'}06{''}$, also named UCAC4 640-050794, CRTS J154915.7+375506, ATO J237.3156+37.9183)
was found as an EW-type eclipsing binary with a period of 0.377998 days and mean $V_{mag}=13.88\,mag$ in the Catalina Surveys Data Release-1 (CSDR1)\citep{drakeCatalinaSurveysPeriodic2014a}.
In 2017, it was included in the Czech Variable Star Catalogue\citep{skarkaCzeVCzechVariable2017}, whereafter it was observed by the Catalina Real-Time Transient Survey (CRTS)\citep{2017MNRAS.465.4678M}
with a light curve displayed and identified as an EB-type binary. This target was observed by the AZT-22 telescope of the Maidanak Astronomical Observatory (MAO), and multicolor light
curves were obtained. It was also observed by the Large sky Area Multi-Object fiber Spectroscopic Telescope (\citep{2012RAA....12.1243L,2012RAA....12..723Z}) in 14th, April 2017, and the
atmospheric parameters of the system were determined. The primary effective temperature $T_1=6929.5\,K$ was adopted in our study. Our results show that this target is a marginal contact binary
system with an extremely low mass ratio of 0.138, which means the mass ratio reversal occurred. The O-C curve analysis demonstrates that a third body may exist.

\section{Photometric and Spectral Observations}
\subsection{Photometric Data}

\begin{figure}
  \includegraphics[width=\columnwidth]{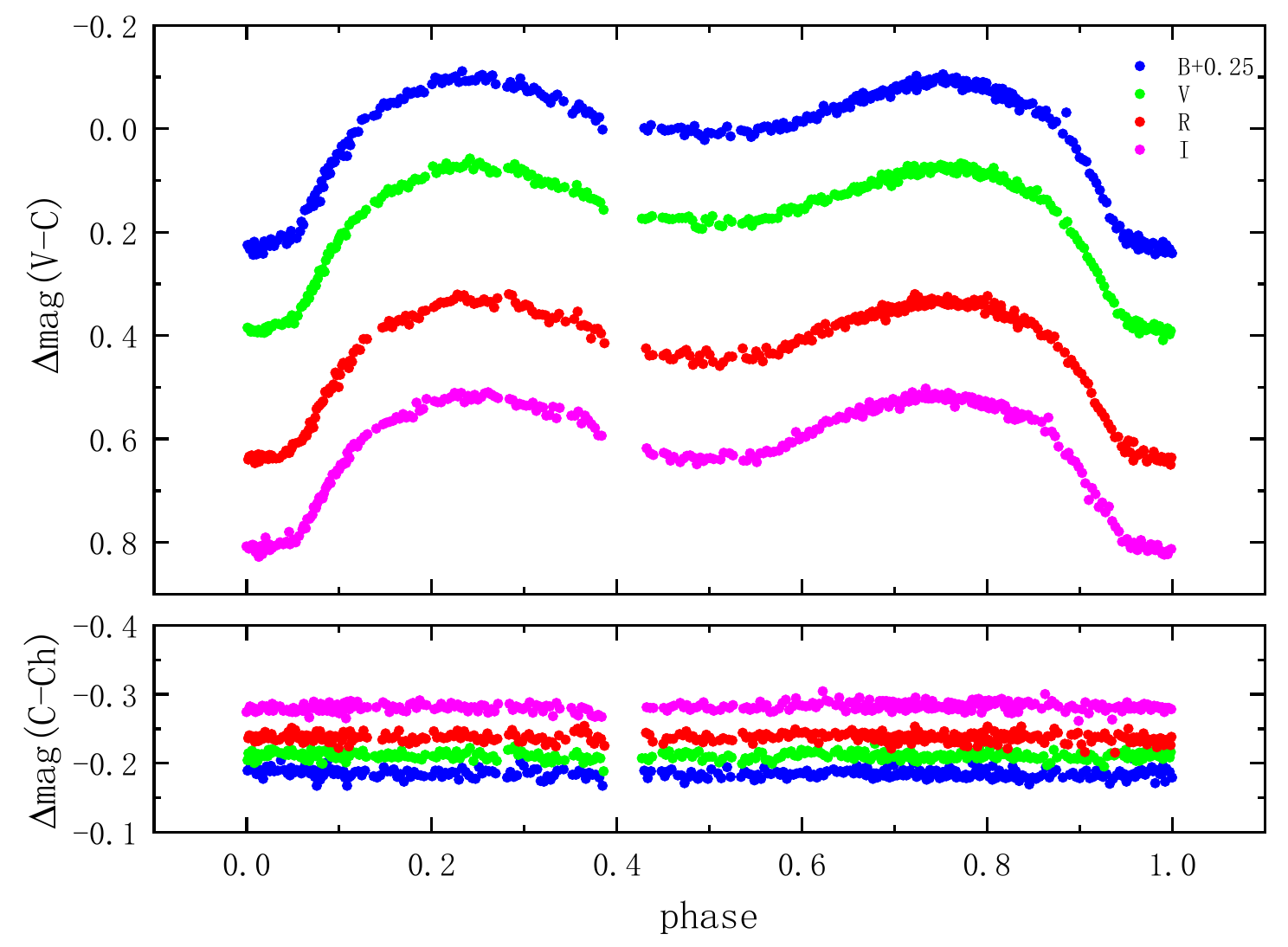}
  \caption{The light curves of CSS\_J154915.7+375506 in the Bessell B, V, $R$ and $I$ bands were observed by the 1.5 m AZT-22 telescope. The magnitude difference of the
   comparison star minus the check star (C-Ch) is shown in the bottom panel.}
  \label{fig:lightcurve}
\end{figure}

New photometric observations of the variable star CSS\_J154915.7+375506 were carried out by the 1.5-m telescope AZT-22 on February 23, April 23, and June 23, 2023, at the  Maidanak
Astronomical Observatory (MAO) of the Ulugh Beg Astronomical Institute of the Uzbekistan Academy of Science \citep{Ehgamberdiev2018}.
The optical system of the AZT-22 telescope is Ritchey-Chretien, and the focal length is 11560 mm \citep{artamonov1998first}. The Andor iKon-XL
(XL-EA07-DS) ($4096 (H) \times 4108 (V)$) CCD camera was used as the receiver. The read-out noise and amplification coefficients of the CCD camera are 6.5ADUs and 1.55 e-, the
effective field of view is $18.1' \times 18.1'$ arcmin, and the pixel scale is 0.268$'$ arcsec per pixel.
In combination with the CCD detector, these provided a photometric system close to the standard Bessell UBVRI system \citep{MyungShin2010JKAS}.
Image acquisition was done with MaxIm DL 6 version. The observations were using the BVRI filters of the Bessell photometric system at 30, 20, 10 and 10 second
exposure times, respectively. Figure~\ref{fig:lightcurve} shows the EB multicolor light curves, and the standard deviation of the comparison minus the check are 0.00683, 0.00524, 0.00622
and 0.00582 for B, V, R and I, respectively. The coordinates of the comparison and check stars are listed in Table~\ref{tab:coordinates}. In the light curves, the difference between the primary
minimum and the secondary is close to 0.2 mag, with the flat primary minimum, this system should be a total eclipse binary.

\begin{table}
  \caption{The coordinates of CSS\_J154915.7+375506, the comparison star and the check star.}
  \label{tab:coordinates}
  \begin{tabular*}{\columnwidth}{@{}l@{\hspace*{13pt}}l@{\hspace*{13pt}}l@{}}
   \hline
   Stars & $\alpha_{2000}$ & $\delta_{2000}$\\

   \hline
   CSS\_J154915.7+375506 & $15^h49^m15^s.76$ & $+37^{\circ}55''06'.01$\\[2pt]
   The comparison & $15^h49^m08^s.88$ & $+37^{\circ}50''21'.19$\\[2pt]
   The check & $15^h49^m09^s.79$ & $+37^{\circ}55''18'.64$\\[2pt]
   \hline
  \end{tabular*}
\end{table}

\subsection{LAMOST Spectrum}
\begin{figure}
  \includegraphics[width=\columnwidth]{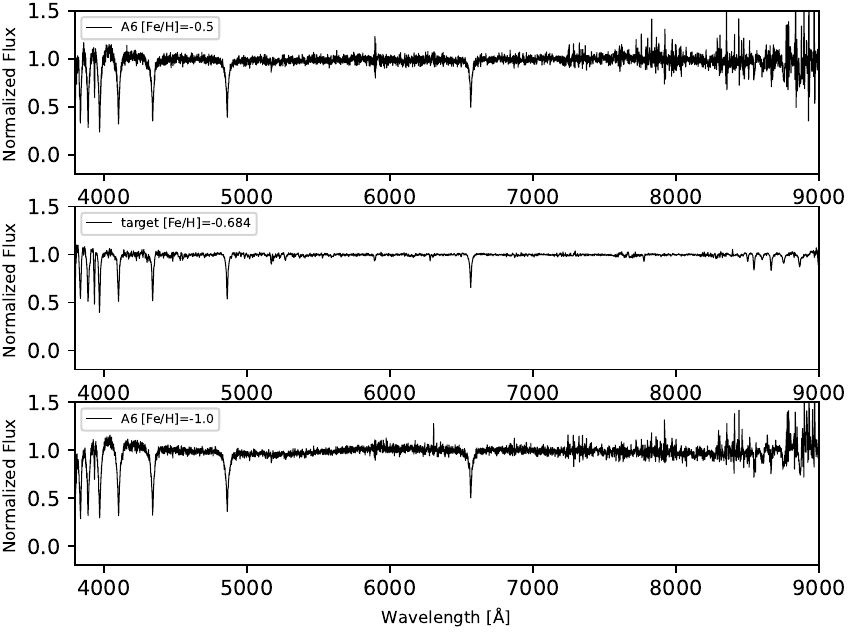}
  \caption{Spectra figures. Upper panel: A normalized A6 spectral template with [Fe/H] = -0.5. Middle panel: the normalized spectrum of CSS\_J154915.7+375506. Lower panel: A normalized A6 spectral template with [Fe/H] = -1.0.}
  \label{fig:LAMOST spectrum}
\end{figure}

LAMOST is a reflective Schmidt telescope with a 4-meter aperture and a 5-degree field of view, which is located at the Xinglong Observatory in Hebei, China.
With 4000 fibers placed on the focal plane, LAMOST can obtain 4000 spectra simultaneously. Meanwhile, stellar atmospheric parameters can be obtained automatically
by the LAMOST stellar parameter pipeline (LASP) \citep{wuAutomaticDeterminationStellar2011,wuAutomaticStellarSpectral2014,luoFirstDataRelease2015}.
The low-resolution spectrum of LAMOST is $R~1800$ and the wavelength range is 3700-9100\AA. We found one low-resolution spectrum from the LAMOST DATA RELEASE 9 and obtained the atmospheric
parameters as spectral type: A6, surface effective temperature: $6929.51\,K$, surface gravity: $4.259\,dex$ and the metallicity: -0.684. The normalized spectrum is shown in
Figure~\ref{fig:LAMOST spectrum}, in which located in the upper panel and lower panel are A6 type spectral template\citep{2017ApJS..230...16K} for comparison.

\section{Light curve Analysis}
\begin{figure}
  \includegraphics[width=\columnwidth]{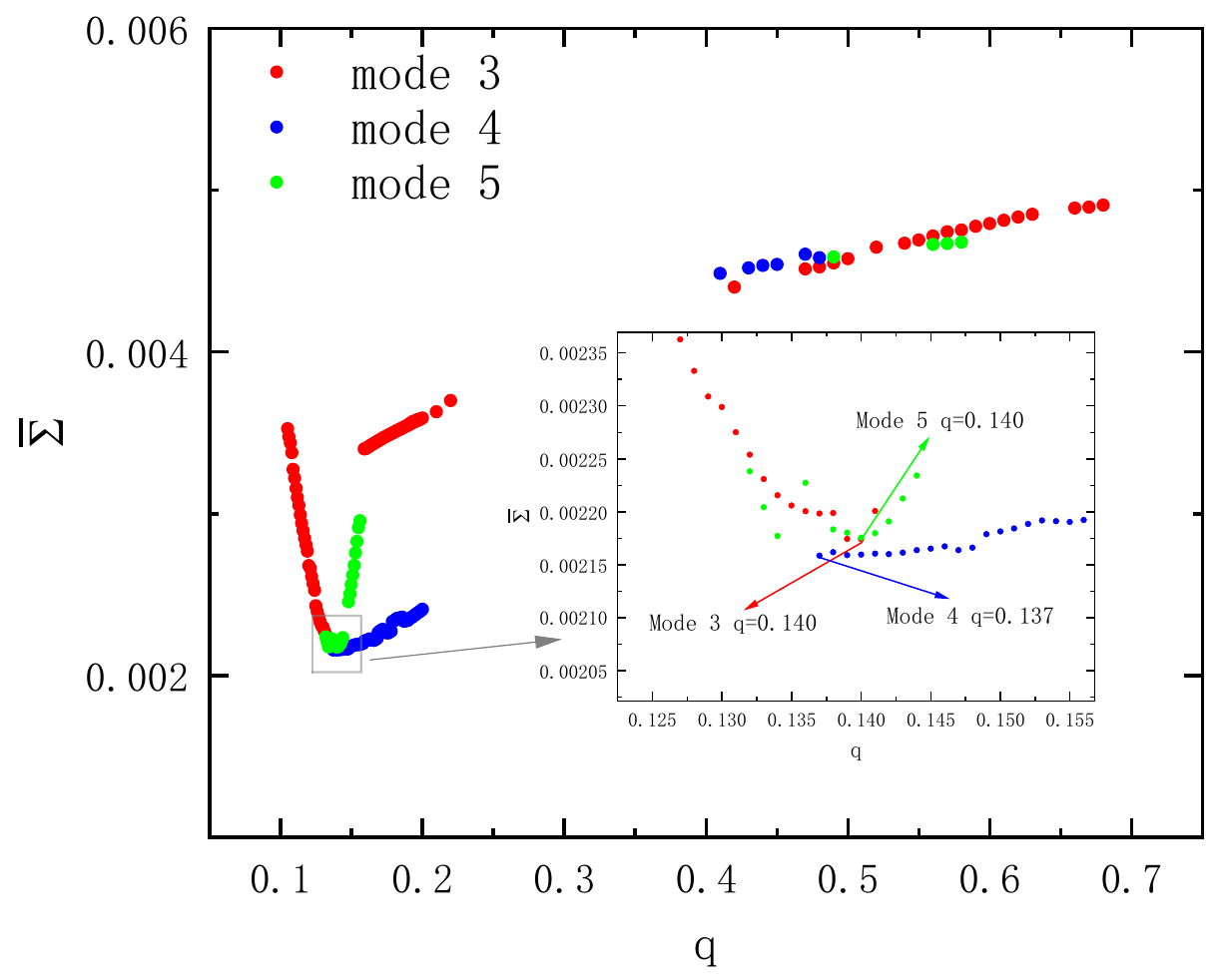}
  \caption{The relationship between $q$ and the mean weighted residuals $\overline{\Sigma}$} for Mode 4 and Mode 5.
  \label{fig:q_search}
\end{figure}

\begin{figure*}  
  \centering
  \includegraphics[width=\linewidth]{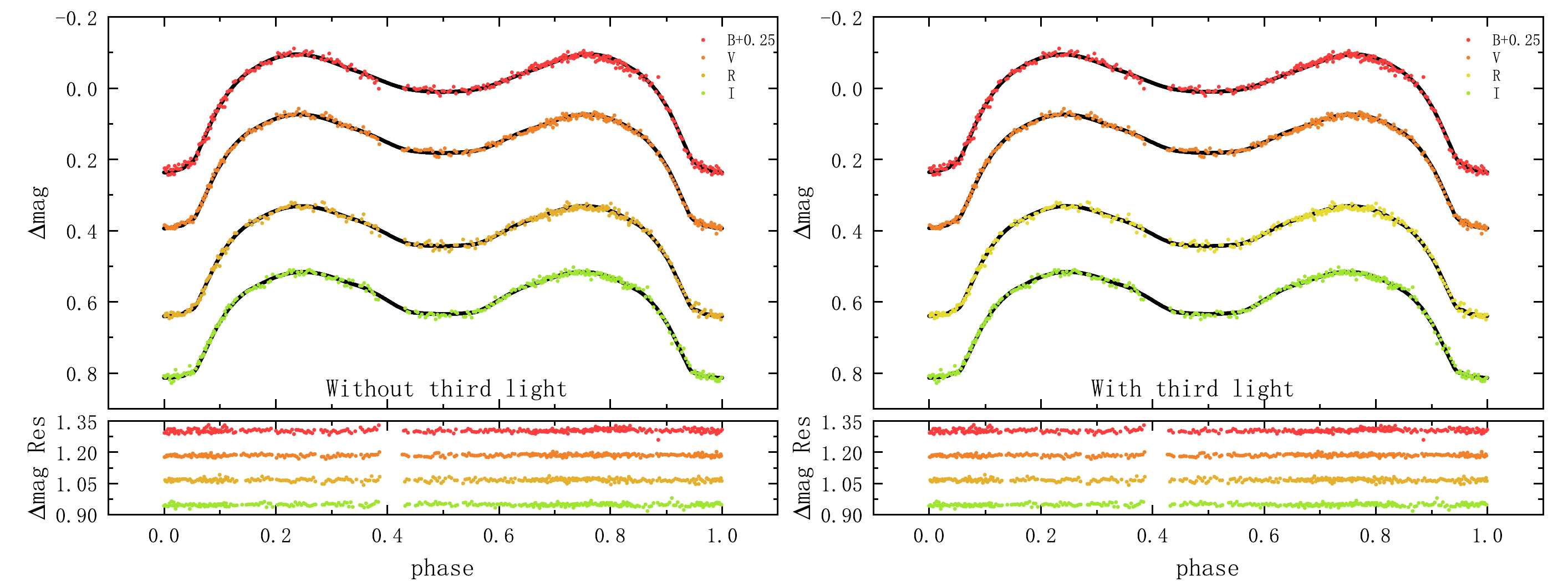}
  \caption{Multiband observational and theoretical light curves of CSS\_J154915.7+375506 in B, V, $R$ and $I$ bands from MAO AZT-22. The left-upper panel is the light curve without $L_3$ while
   the right-upper panel is the light curve with $L_3$. In the bottom panel, the residuals of the observations minus the theoretical light curves are shown. The residuals are vertical moved to
   display well.}
  \label{fig:Observation and fitted}
\end{figure*}

The Wilson-Devinney\citep{1971ApJ...166..605W,1990ApJ...356..613W, Wilson_2012, Wilson_2014} code was applied to analyze the multicolor light curves. The surface effective temperature of
the primary component was set to $T_1$ = 6930 K according to the stellar atmospheric parameters of LAMOST. The gravity-darkening coefficients $g_1$ = 0.32 and $g_2$ = 0.32\citep{1967ZA.....65...89L}, the
bolometric albedo $A_1$ = 0.5 and $A_2$ = 0.5 \citep{rucinski1969proximity} were adopted. We tried to adopt mode 2 (detached binary), mode 3 (overcontact binary), mode 4 (semidetached binary with star 1 nearly
filling its Roche lobe) and mode 5 (semidetached binary with star 2 nearly filling its Roche lobe) to fit the light curves but mode 2 failed to converge. We used the q-search method to
derive initial mass ratio. For mode 3, 4 and 5, convergent solutions were obtained when the mass ratio was fixed as a series of values from 0.1 to 1 with a step of 0.001.
The minimum values of $\Sigma$ were achieved at similar q values: 0.140, 0.137 and 0.140, respectively, which indicates the reliability of the mass ratio. The q-search diagram is
shown in Figure~\ref{fig:q_search}. After that, we set the q with the value 0.137 and rerun to run the WD code for the three modes. The other adjustable parameters of mode 3 are the
orbital inclination, $i$, the mean surface temperature of the secondary component, $T_2$, the dimensionless potential of the primary, $\Omega_1$ (with $\Omega_2=\Omega_1$), the
band-pass luminosity of the primary, $L_1$, and the third light, $L_3$. For mode 4, the only difference with mode 3 is the adjustable dimensionless surface potential is of the
secondary, $\Omega_2$. As for mode 5, the adjustable parameters are the same as that of mode 3 (with $\Omega_1\neq\Omega_2$). All the convergent solutions are listed in
Table~\ref{tab:photometric solutions}. The $\Sigma$ can be noticed in the last row of the table, and no large difference exists; as an example, the fitted light curves of mode 3 with and
without $L_3$ are shown in Figure~\ref{fig:Observation and fitted}. It can be seen from Figure~\ref{fig:Observation and fitted} that the fits with $L_3$ look almost identical to the fits
without $L_3$ but smaller residuals can be noticed from the Table~\ref{tab:photometric solutions}. So CSS\_J154915.7+375506 may be a marginal contact binary with an extremely low mass ratio, with the
low contribution of third light, a faint third companion could exist.

\begin{table*}
  \centering
  \caption{Photometric solutions of CSS\_J154915.7+375506 obtained by using the W-D code. The units of most parameters are dimensionless, except those already mentioned.}
  \label{tab:photometric solutions}
  \begin{tabularx}{\textwidth}{p{3cm}*{6}{>{\centering\arraybackslash}X}}
    \hline
                             & \multicolumn{2}{c}{Mode 3}           & \multicolumn{2}{c}{Mode 4}         & \multicolumn{2}{c}{Mode 5} \\
    Parameters                      & Without $L_3$  & With $L_3$       & Without $L_3$ & With $L_3$         & Without $L_3$ & With $L_3$\\
    \hline
    $g_1=g_2$                       & 0.32           & 0.32             & 0.32          & 0.32               & 0.32          & 0.32               \\
    $A_1=A_2$                       & 0.5            & 0.5              & 0.5           & 0.5                & 0.5           & 0.5                \\
    $q(M_2/M_1)$                    & 0.13875(55)    & 0.13824(91)      & 0.13828(46)   & 0.13828(89)        & 0.13729(28)   & 0.13696(61)        \\
    $i(^{\circ})$                   & 86.153(97)     & 86.151(97)       & 85.796(95)    & 85.796(98)         & 87.17(12)     & 87.02(11)          \\
    $T_1(K)$                        & 6930           & 6930             & 6930          & 6930               & 6930          & 6930               \\
    $T_2(K) $                       & 4209(33)       & 4210(33)         & 4184(30)      & 4184(31)           & 4225(30)      & 4223(32)           \\
    $\Omega_1$                      & 2.0720(22)     & 2.0706(24)       & 2.07088       & 2.07088            & 2.07093(36)   & 2.07007(43)        \\
    $\Omega_2$                      & 2.0720(22)     & 2.0706(24)       & 2.0714(21)    & 2.0714(21)         & 2.06814       & 2.06722            \\
    $L_1/(L_1 + L_2)_B$             & 0.9951413(26)  & 0.995146(35)     & 0.9954392(21) & 0.995439(39)       & 0.9949932(25) & 0.995026(22)       \\
    $L_1/(L_1 + L_2)_V$             & 0.9892739(82)  & 0.989289(86)     & 0.9898197(72) & 0.989820(97)       & 0.9890236(81) & 0.989085(59)       \\
    $L_1/(L_1 + L_2)_R$             & 0.982783(16)   & 0.98281(15)      & 0.983541(15)  & 0.98354(17)        & 0.982461(16)  & 0.98255(11)        \\
    $L_1/(L_1 + L_2)_I$             & 0.974829(26)   & 0.97488(24)      & 0.975723(25)  & 0.97572(27)        & 0.974491(26)  & 0.97460(18)        \\
    $L_3/(L_1 + L_2 + L_3)_B$       & --             & 0.0131(65)       & --            & 0.0126(75)         & --            & 0.0113(42)         \\
    $L_3/(L_1 + L_2 + L_3)_V$       & --             & 0.010711(86)     & --            & 0.0018(86)         & --            & 0.0003(51)         \\
    $L_3/(L_1 + L_2 + L_3)_R$       & --             & 0.00048(838)     & --            & 0.0011(93)         & --            & 0.0000(61)         \\
    $L_3/(L_1 + L_2 + L_3)_I$       & --             & $9.4\times10^{-5}(9.2\times10^{-3})$& --& 0.0019(102) & --            & 0.0000(70)         \\
    $r_1(pole)$                     & 0.51311(44)    & 0.51329(45)      & 0.51338(23)   & 0.51328(44)        & 0.51316(10)   & 0.51328(16)        \\
    $r_1(side)$                     & 0.56504(64)    & 0.5653(065)      & 0.56543(32)   & 0.56529(62)        & 0.56505(16)   & 0.56521(27)        \\
    $r_1(back)$                     & 0.58524(67)    & 0.58551(67)      & 0.58566(3)    & 0.58552(57)        & 0.58504(21)   & 0.58518(40)        \\
    $r_2(pole)$                     & 0.2175(21)     & 0.2176(3)        & 0.2172(18)    & 0.2175(27)         & 0.21698(13)   & 0.21684(28)        \\
    $r_2(side)$                     & 0.2092(18)     & 0.2093(25)       & 0.209(16)     & 0.2092(23)         & 0.20869(13)   & 0.20856(27)        \\
    $r_2(back)$                     & 0.2492(40)     & 0.2495(58)       & 0.2488(35)    & 0.2492(53)         & 0.24868(14)   & 0.24853(29)        \\
    $fill-out factor$               & 0.00231(0.02415)&0.00228(0.026452)& --            & --                 & --            & --                 \\
    $\Sigma$                        & 0.00204        & 0.00169          & 0.00203       & 0.00169            & 0.00203       & 0.00170            \\
    \hline
  \end{tabularx}
\end{table*}

\section{Orbital Period Investigation}

\begin{table}
  \caption{The light minima of CSS\_J154915.7+375506.}
  \label{tab:light minima}
  \begin{tabular*}{\columnwidth}{@{}l@{\hspace*{13pt}}l@{\hspace*{13pt}}l@{\hspace*{13pt}}l@{\hspace*{13pt}}l@{}}
   \hline
   Eclipse timings & Error & E & O-C & Reference\\
   (HJD) & ($\pm$ d) &  & (d) & \\
   \hline
   2457208.72  & 0.001432612 & 0     & 0.00000000  & ASAS-SN  \\[2pt]
   2458005.934 & 0.001436392 & 2109  & 0.01637465  & ASAS-SN  \\[2pt]
   2454143.537 & 0.000506517 & -8109 & 0.00223421  & CRTS     \\[2pt]
   2455615.454 & 0.000971455 & -4215 & -0.00444866 & CRTS     \\[2pt]
   2458269.408 & 0.000374218 & 2806  & 0.02509633  & ZTF      \\[2pt]
   2458630.022 & 0.000442258 & 3760  & 0.02942852  & ZTF      \\[2pt]
   2459015.204 & 0.000351538 & 4779  & 0.03178869  & ZTF      \\[2pt]
   2459363.34  & 0.000408238 & 5700  & 0.03150496  & ZTF      \\[2pt]
   2459743.98  & 0.000782456 & 6707  & 0.0274099   & ZTF      \\[2pt]
   2460161.285 & 0.0002675   & 7811  & 0.02224803  & AZT-22   \\[2pt]
   2458962.284 & 0.000287278 & 4639  & 0.03109575  & TESS s24 \\[2pt]
   2458975.892 & 0.000238139 & 4675  & 0.03143317  & TESS s24 \\[2pt]
   2459669.139 & 0.000487617 & 6509  & 0.03005764  & TESS s50 \\[2pt]
   2459675.565 & 0.000449818 & 6526  & 0.03010509  & TESS s50 \\[2pt] 
   2459682.747 & 0.000540537 & 6545  & 0.0295658   & TESS s50 \\[2pt]
   2459688.795 & 0.000389338 & 6561  & 0.0295849   & TESS s50 \\[2pt]
   2459695.976 & 0.000287278 & 6580  & 0.02904642  & TESS s51 \\[2pt]
   2459702.024 & 0.000241919 & 6596  & 0.02900456  & TESS s51 \\[2pt]
   2459708.828 & 0.000351538 & 6614  & 0.02884077  & TESS s51 \\[2pt]
   2459714.876 & 0.000268379 & 6630  & 0.02880401  & TESS s51 \\[2pt]
   \hline
  \end{tabular*}
\end{table}

\begin{figure*}
  \centering
  \includegraphics[width=\linewidth]{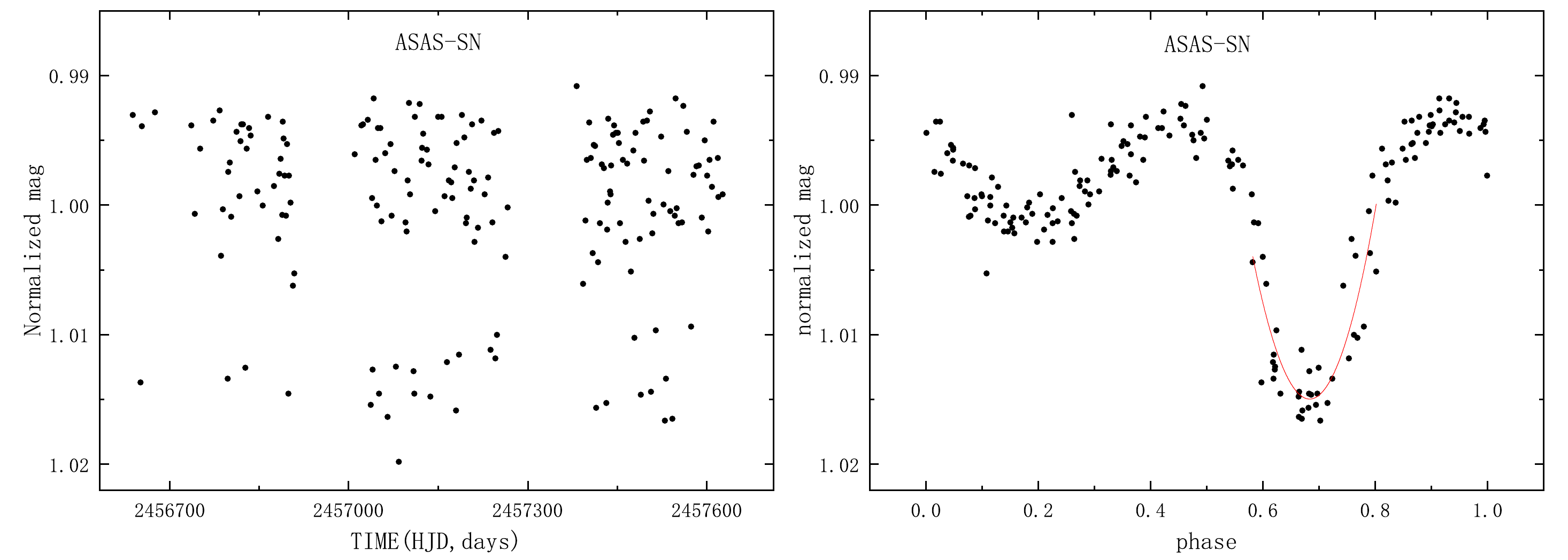}
  \caption{Left panel: ASAS-SN photometric data in V band. Right panel: Light curve converted to one phase and fitted the light minima phase with a parabola function.}
  \label{fig:ASAS-SN phase}
\end{figure*}

\begin{table}
  \caption{Orbital parameters of the third body in CSS\_J154915.7+375506.}
  \label{tab:third body pars}
  \begin{tabular*}{\columnwidth}{@{}l@{\hspace*{6pt}}l@{\hspace*{6pt}}l@{}}
    \hline
    \hline
    Parameter & Value & Unit\\
    \textbf{Case A: linear term plus}                 &                                 &                \\[2pt]
    \textbf{cyclical variation($\beta=0$)}            &                                 &                \\[2pt]
    Revised epoch, $T_0$                              & 2457208.7277($\pm$0.0007)       & HJD            \\[2pt]
    Revised period, $P_0$                             & 0.37800020($\pm$0.00000010)     & days           \\[2pt]
    Eccentricity, $e_3$                               & 0 (assumed)                     & --             \\[2pt]
    Long-term change of the orb-\\
    \phantom{{}={}}ital period, $\beta$               & 0 (assumed)                     & d cycle$^{-1}$ \\[2pt]
    Orbital period, $P_3$                             & 12.48($\pm$0.25)                & yr             \\[2pt]
    Amplitude, A                                      & 0.0142($\pm$0.0006)             & days           \\[2pt]
    Projected semi-major axis,$a_{12}sin(i_3)$        & 2.45($\pm$0.10)                 & au             \\[2pt]
    Mass function, $f(m)$                             & 0.10($\pm$0.01)                 & $M_{\odot}$    \\[2pt]
    Mass, $M_{3min}$                                  & 1.02($\pm$0.05)                 & $M_{\odot}$    \\[2pt]
    Sum of the squares of the residuals, $\Sigma$     & $1.66\times 10^{-5}$            & --             \\[2pt]
    \hline

    \textbf{Case B: quadratic term plus}              &                                 &                \\[2pt]
    \textbf{cyclical variation($\beta \neq 0$)}       &                                 &                \\[2pt]
    Revised epoch, $T_0$                              & 2457208.7335($\pm$0.0017)       & HJD            \\[2pt]
    Revised period, $P_0$                             & 0.37798160($\pm$0.00000030)     & days           \\[2pt]
    Eccentricity, $e_3$                               & 0 (assumed)                     & --             \\[2pt]
    Long-term change of the orb-\\
    \phantom{{}={}}ital period, $\beta$               & $-3.25(\pm0.89) \times 10^{-10}$   & d cycle$^{-1}$ \\[2pt]
    Orbital period, $P_3$                             & 14.89($\pm$0.96)                & yr             \\[2pt]
    Amplitude, A                                      & 0.0145($\pm$0.0010)             & days           \\[2pt]
    Projected semi-major axis,$a_{12}sin(i_3)$        & 2.51($\pm$0.17)                 & au             \\[2pt]
    Mass function, $f(m)$                             & 0.07($\pm$0.02)                 & $M_{\odot}$    \\[2pt]
    Mass, $M_{3min}$                                  & 0.91($\pm$0.06)                 & $M_{\odot}$    \\[2pt]
    Sum of the squares of the residuals, $\Sigma$     & $2.98\times 10^{-5}$            & --             \\[2pt]
    \hline

    \textbf{Case C: linear term plus} & & \\
    \textbf{eccentric orbit($\beta=0$)} & & \\
    Revised epoch, $T_0$                              & 2457208.7237($\pm$0.0030)       & HJD            \\[2pt]
    Revised period, $P_0$                             & 0.37800028($\pm$0.00000005)     & days           \\[2pt]
    Eccentricity, $e_3$                               & 0.40($\pm$0.15)                 & --             \\[2pt]
    Long-term change of the orb-\\
    \phantom{{}={}}ital period, $\beta$               & 0 (assumed)                     & d cycle$^{-1}$ \\[2pt]
    Orbital period, $P_3$                             & 12.37($\pm$0.11)                & yr             \\[2pt]
    Amplitude, A                                      & 0.0175($\pm$0.0029)             & days           \\[2pt]
    Longitude of the periastron\\
    \phantom{{}={}}passage, $\omega_3$                & 260.02($\pm$8.85)               & degree         \\[2pt]
    Time of periastron passage, $T_3$                 & 2456409.0183($\pm$120.5700)     & days           \\[2pt]
    Projected semi-major axis,$a_{12}sin(i_3)$        & 3.03($\pm$0.50)                 & au             \\[2pt]
    Mass function, $f(m)$                             & 0.18($\pm$0.09)                 & $M_{\odot}$    \\[2pt]
    Mass, $M_{3min}$                                  & 1.36($\pm$0.19)                 & $M_{\odot}$    \\[2pt]
    Sum of the squares of the residuals, $\Sigma$     & $1.73\times 10^{-5}$            & --             \\[2pt]
    \hline
   \end{tabular*}
\end{table}

Orbital period variations are common in binary systems which are usually caused by angular momentum loss, mass transfer, third body or magnetic activity. We used
the O-C method to analyze the orbital period variation of this system. To obtain more light minimum times, we tried
to search for more photometric data from the sky survey project and public databases. We collected data from All-Sky Automated Survey for Supernovae (ASAS-SN)\citep{2014ApJ...788...48S,2023MNRAS.519.5271C} database, Catalina Real-Time
Transient Survey (CRTS)\citep{2009ApJ...696..870D} database, The Zwicky Transient Facility (ZTF)\citep{Bellm_2019} and Transiting Exoplanet Survey Satellite (TESS)
\citep{ricker2010transiting} which provided Full Frame Images (FFI) for three sectors with an exposure time of the 1800s, 600s and 600s. We converted data with more than one cycle into a phase
\citep{10.1093/mnras/stab1657,2022ApJ...924...30L} to get the minima by using parabola fitting. Some data of the ASAS-SN is shown in the left panel of Figure.~\ref{fig:ASAS-SN phase}
and the right panel shows the light curve converted to one phase. Finally, we obtained 2 times of minimum from ASAS-SN, 2 times of minimum from CRTS, 5 times of minimum from ZTF and
10 times of minimum from TESS, plus the 1 fitted from AZT-22 data, 20 in total. The times of light minima are listed in the first column of the Table.~\ref{tab:light minima}. The $O-C$
values were calculated with the following linear ephemeris formula:
\begin{equation}
  Min.I(HJD) = 2457208.72017397 + 0^d.377998 \times E
  \label{eq:linear ephemeris}
\end{equation}
\noindent In which the $HJD_0$ and period value was obtained from The International Variable Star Index(VSX)\footnote{\url{https://www.aavso.org/vsx/}}. In O-C curves, it is clear
that a periodic oscillation exists which may be caused by the light time travel effect(LTTE) of a third body\citep{liao2010most}. We use the following general equation to fit
the O-C diagram\citep{irwinDeterminationLightTimeOrbit1952}:

\begin{equation}
  \begin{aligned}
  O-C&=\Delta T_0+\Delta P_0 E + \frac{\beta}{2} E^2 \\
  &\phantom{{}={}}+ A[(1-e^2_3)\frac{sin(\nu + \omega)}{1+e_3cos\nu}+esin\omega] \\
  &=\Delta T_0 + \Delta p_0E+\frac{\beta}{2}E^2 \\
  &\phantom{{}={}}+ A[\sqrt{1-e_3^2}sinE^*cos\omega +cosE^*sin\omega]
  \end{aligned}
  \label{eq:O-C}
\end{equation}

\begin{figure*}
  \includegraphics[width=\linewidth]{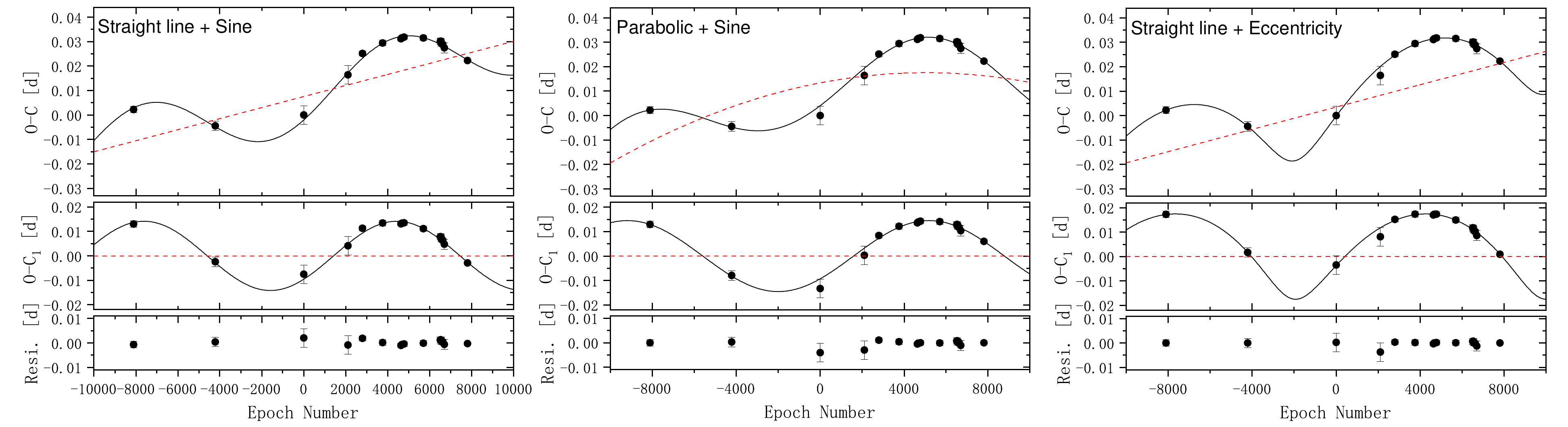}
  \caption{The O-C curve from the linear ephemeris of equation (\ref{eq:linear ephemeris}) are shown in the upper panel, the middle panel shows the fitted results of linear term
   plus cyclical variation, quadratic term plus cyclical variation and linear term plus eccentric orbit and the bottom panel shows the residuals of O-C curve removed all variations.}
  \label{fig:O-C-fit}
\end{figure*}

\noindent The Kepler equation is as follows:

\begin{equation}
  M=E^*-e_3sinE^*=\frac{2\pi}{P_3}(t-T_3)
  \label{eq:Kepler equation}
\end{equation}

\noindent the meaning of each parameter is described very clearly in the paper \citep{liaoOrbitalParametersHierarchical2021}. The corresponding results of Equations\ref{eq:O-C}
and \ref{eq:Kepler equation} are displayed in Table~\ref{tab:third body pars}. Based on the existence of the periodic oscillation, we assumed a circular orbit($e_3=0$), i.e. (1)
Case A, the linear term plus cyclical variation($\beta=0$) and (2) Case B, the quadratic term plus cyclical variation($\beta \neq 0$). In addition, we considered an eccentric orbit
($e_3 \neq 0$) with a linear term, i.e. (3) Case C, the linear term plus eccentric orbit($\beta=0$). By using Equation~\ref{eq:O-C} and \ref{eq:Kepler equation} to fit the O-C cure,
the results of the three cases are listed in Table~\ref{tab:third body pars}. The fitted O-C curve and the residuals for different cases are shown in Figure~\ref{fig:O-C-fit}.
For Case A, the amplitude and period of the cyclical variation are $A=0.0142(\pm0.0006)\,days$ and $P_3=12.48(\pm0.25)\,yr$.
For Case B, the amplitude and period of the cyclical variation are $0.0145(\pm0.0010)\,days$ and $P_3=14.89(\pm0.96)\,yr$ with a long-term period decrease at a rate:
$\beta=-3.25 \times 10^{-10}\,d\,cycle^{-1}$, i.e. $-3.14(\pm0.89) \times 10^{-7}\,days\,yr^{-1}$. We can see different possible O-C curve variations correspond to different
cases, but a common periodic oscillation may imply an existence of a third body, so further investigations are necessary. As for the Case C, the eccentricity of the third body
is $e_3=0.40(\pm0.15)$, the amplitude and period of the cyclical variation are $A=0.0175(\pm0.0029)\,days$ and $P_3=12.37(\pm0.11)\,yr$, respectively. 

\section{Discussion and Conclusion}
\subsection{Absolute parameters}

\begin{table}
  \caption{Estimated absolute parameters of CSS\_J154915.7+375506}
  \label{tab:absolute parameters}
  \begin{tabular*}{\columnwidth}{@{}l@{\hspace*{28pt}}l@{\hspace*{28pt}}l@{\hspace*{28pt}}l@{}}
   \hline
   Parameters   & Value  & Parameters & Value\\
   \hline
   $\varpi$ (mas) & 0.6934(0.014)   &  $A_v$   & 0.035\\
   $m_V$          & 13.776(0.023)   &  $M_v$   &  2.95(0.05)\\
   $M_{bol}$      & 2.88(0.05)      &  $BC_V$  &  -0.07\\
   $M_1$          & 2.05(0.15)      &  $M_2$   &  0.283(0.02)\\
   $R_1$          & 1.62(0.04)      &  $R_2$   &  0.656(0.016)\\
   $L_1$          & 5.43(0.27)      &  $L_2$   &  0.121(0.006)\\
   semi major axis& 2.92(0.07)      &          &              \\
   \hline
  \end{tabular*}
\end{table}

The orbital period analysis and multicolor CCD photometric solutions of CSS\_J154915.7+375506 are presented for the first time. With the newly observed multicolor light
curves, good photometric solutions on mode 3, mode 4 and mode 5 were obtained by using W-D code. Although the radial velocity curve is missing, it is a total
eclipse binary system, and all solutions show similar mass ratios, so the results are reliable. The solutions are shown to us in Table~\ref{tab:photometric solutions} that all the
convergent solutions are with very close parameter values and similar residuals, and all solutions with $L_3$ have smaller residuals, so we will take the average values of
the results with $L_3$ for the following discussions.

Combined with the photometric solutions and parallax, the absolute parameters can be estimated by the method described by \citep{2021AJ....162...13L} and \citep{xuV0644SerActive2022}.
The parallax $\varpi$ was given by Gaia Dr3 as 0.6934 mas, so the distance was estimated as about 1442 pc, which was calculated by the following formula:

\begin{equation}
  (m-M)_v=10-5lg\varpi+A_v
\end{equation}

\noindent where the extinction $A_v$ is 0.035 mag (obtained from the NASA/IPAC Extragalactic Database\footnote{\url{https://ned.ipac.caltech.edu/extinction_calculator}}.), and the
apparent magnitude $m_v$ is 13.776 mag, the maximum brightness in the V band from ASAS-SN data\citep{2014ApJ...788...48S,2023MNRAS.519.5271C}. Then with the bolometric correction
$BC_v=$ -0.07 mag\citep{2011yCat..21930001W}, we can obtain its bolometric absolute magnitude $M_v$ with this equation:

\begin{equation}
  M_{bol}=M_v+BC_v
\end{equation}

\noindent hence, the total luminosity $L_t$ of the binary system is derived by:

\begin{equation}
  L_T=L_1+L_2=10^{(M_{bol}-4.74)/-2.5}
\end{equation}

\noindent And with the Stefan Boltzmann's law, we have:

\begin{equation}
  L_t/L_{\odot}=(ar_1)^2\times(\frac{T_1}{T_{\odot}})^4+(ar_2)^2\times(\frac{T_2}{T_{\odot}})^4
\end{equation}

\noindent where $T_{\odot}$ represents the effective temperature of the sun, with a value of 5772 K. $r_1$ and $r_2$ represent the relative radii of the star 1 and star 2, respectively,
which can be obtained from the photometric solutions, just like $r_i=(r_{ipole}*r_{iside}*r_{iback})^{1/3}$. The semi-major axis $a$ is obtained as about $2.92\,R_{\odot}$. Combined with the
mass ratio and the Kepler's third law,

\begin{equation}
  \frac{a^3}{P^2}=74.5(M_1+M_2),
\end{equation}

\begin{equation}
  q=\frac{M_2}{M_1}
\end{equation}

\noindent the mass of the primary component $M_1$ and the mass of the secondary component $M_2$ were calculated about 2.05 $M_{\odot}$ and 0.283 $M_{\odot}$, respectively.
All the estimated absolute parameters are tabulated in Table~\ref{tab:absolute parameters}.

\subsection{Orbital period variations}
In the present paper, the orbital period changes are studied for the first time by analyzing a total of 20 times of minima with time span of 16 years. In Case B, a long-term period decrease
and periodic oscillation are revealed with $\beta=-3.14(\pm0.89) \times 10^{-7}\,days\,yr^{-1}$. According to the period variation equation:

\begin{equation}
  \frac{\dot{P}}{P}=-3(\frac{\dot{M_1}}{M_1}+\frac{\dot{M_2}}{M_2}),
\end{equation}
\noindent we can get the rate of mass transfer from the primary to the secondary is $\dot{M_1}=-9.1(\pm2.8) \times 10^{-8} M_{\odot} \, yr^{-1}$. Without more long-term observations, we cannot
determine whether the period decrease exists or not. In general, all cases of O-C curve analysis show evidences of the periodic oscillation. Considering that the target is an A-type marginal
contact binary and photometric solutions with third lights from symmetrical light curves were obtained, the oscillations are mostly caused by the LTTE of a third body. In order to do further
study on the period variation, we use the following formula to estimate the possible parameters of the third body:

\begin{equation}
  f(m)=\frac{(M_3sini_3)^3}{(M_1+M_2+M_3)^2}=\frac{4\pi^2}{GP_3^2}\times (a_{12}sini_3)^3,
  \label{eq:mass function}
\end{equation}

\noindent the mass function of this system is $f(m)=0.10(\pm0.01)\,M_{\odot}$ for Case A, $f(m)=0.07(\pm0.02)\,M_{\odot}$ for Case B and $f(m)=0.18(\pm0.09)\,M_{\odot}$ for Case C.
By assuming the orbital inclination $i_3=90^{\circ}$, we can estimate the minimum mass of the third body are $M_3=1.02(\pm0.05)\,M_{\odot}$, $M_3=0.91(\pm0.06)\,M_{\odot}$ and $M_3=1.36(\pm0.19)\,M_{\odot}$ respectively.
However, considering the extremely low contribution of the third light, this third body may be not in its main-sequence stage, perhaps an unseen compact object.
Some statistics indicate that third bodies are common in contact binaries\citep{2006A&A...450..681T,2006AJ....131.2986P,2006AJ....132..650D,2007AJ....134.2353R}, and the existence of a third body
may play an important role in the binary's evolution. Core fragmentation is thought to be the way to form binary system\citep{1986ApJS...62..519B,1995MNRAS.277..362B}. However, binaries
yielded by this mechanism correspond to initial separations range of 10 to 1000 au. One way of forming closer binaries is dynamical interactions with other nearby stars\citep{2002MNRAS.336..705B}.
Dynamical interactions often result in the replacement of the lower-mass binary component with a higher-mass third body. Consequently, the original system undergoes a process where the
lower-mass component is ejected to a wider orbit, forming a hierarchical triple system. Through the passage of time, these processes have a tendency to give rise to binary systems featuring closely
spaced components and almost equivalent masses\citep{2002MNRAS.336..705B,2013AJ....145....3G}. Based on our system, the mass of the third body is greater than that of the secondary component,
indicating that CSS\_J154915.7+375506 is the original one and was not replaced by the third body. Therefore, the third carried some dynamical characteristics during binary star formation.

\subsection{Discussion}

\begin{figure}
  \includegraphics[width=\columnwidth]{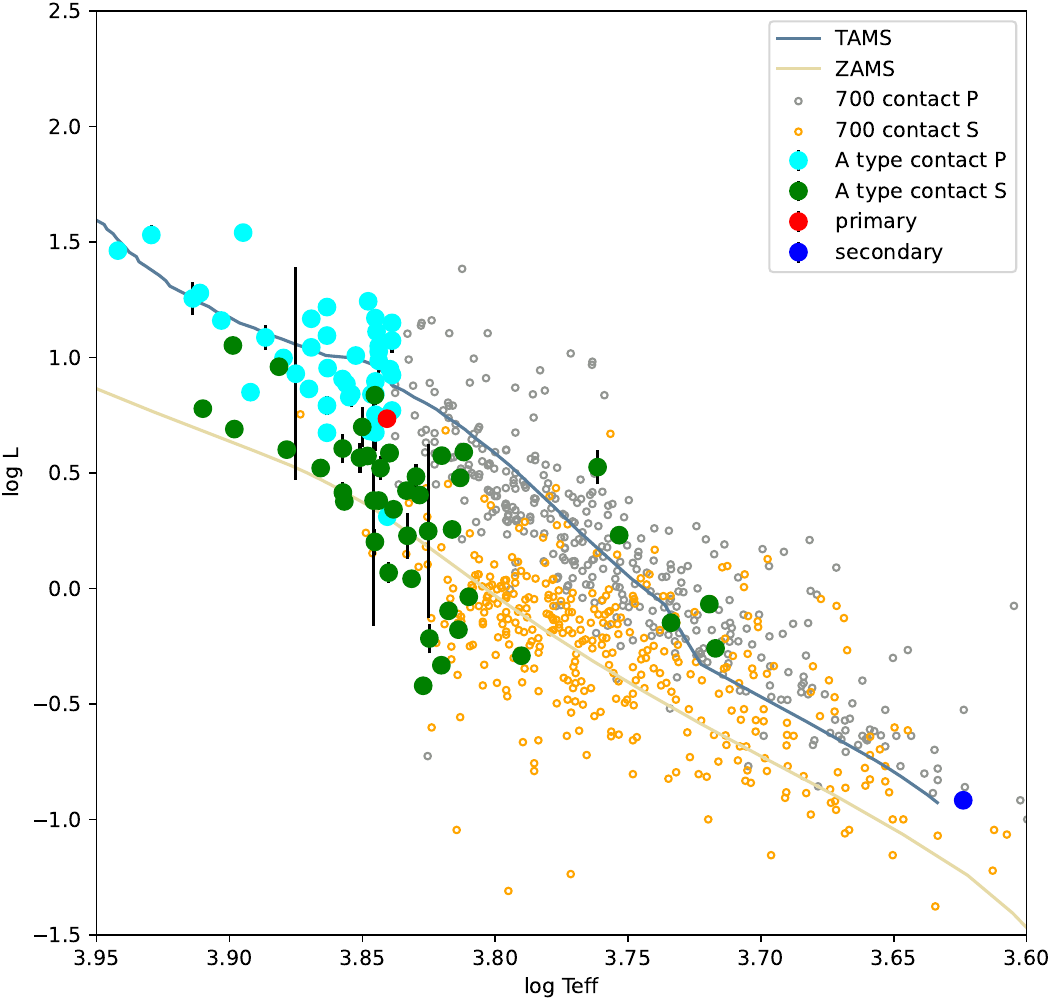}
  \caption{The positions of the primary and secondary component of CSS\_J154915.7+375506 in the H-R diagram.}
  \label{fig:HR diagram}
\end{figure}

Hertzsprung-Russell diagram with [Fe/H] = -0.75 was displayed in Figure~\ref{fig:HR diagram}. The primary and secondary of this system are plotted in this figure. Besides, we plotted 700
contact binaries\citep{2021ApJS..254...10L} and selected spectral A type contact binaries by setting the primary temperature range $6900\,K < T_1 < 9000\,K$. From this diagram, normal contact
binaries' primary components are usually more evolved, while the secondary is not. As for A type contact binaries, most of the primaries are evolved to near the TAMS(terminal-age main-sequence), and
the secondaries just went through the ZAMS(zero-age main-sequence) \citep{2016ApJS..222....8D,2016ApJ...823..102C}. CSS\_J154915.7+375506 is different, the primary component is in main-sequence near the TAMS, while the secondary seems to just
deviate from the TAMS. Under this situation: the extremely low mass ratio of 0.138 with the mass of the secondary of $0.283\,M_{\odot}$, which is much less than the primary component.
The secondary should not have evolved so quickly on its own\citep{clayton1983principles,1971ARA&A...9..183P}, so it must be caused by the interaction between the two components.

Now let's consider the aspect of mutual influence between the binary components. On the one hand, the secondary star has a much smaller mass than the primary
star, but it is more evolved, indicating that the secondary star was once the primary star. It evolved faster and transferred a large amount of mass to the current primary star\citep{crawford1955subgiant}.
Its characteristic of marginal contact indicates that this system has just experienced a mass ratio reversal and is evolving towards the next stage.

On the other hand, the current photometric solutions indicate that the mass of the secondary star is $0.283\,M_{\odot}$, but the temperature is higher than 4000 K. This suggests that the
secondary star, after evolving and expanding, has shed its outer material and exposed part of its core, which is why its current temperature is relatively high. According to the O-C curve
analysis, if the long-term period decreasing really exists, the system will evolve to be a contact binary.

\subsection{Conclusion}
In conclusion, the distinct characteristics of CSS\_J154915.7+375506, including its exceptionally low mass ratio and the evolutionary status of its secondary component, together with its high
temperature relative to its mass, suggest that the system has undergone significant mass transfer, likely experienced a mass ratio reversal event. Furthermore, considering the evidence
from current photometric solutions and the potential for long-term period decrease based on O-C curve analysis, the system may evolve towards becoming a contact binary. Therefore, CSS\_J154915.7+375506
serves as an intriguing subject for investigating the formation processes and evolutionary pathways of binary systems.

\section{DATA AVAILABILITY}
The data underlying this article will be uploaded to the VizieR data base (CDS).

\section*{Acknowledgements}

\addcontentsline{toc}{section}{Acknowledgements}

This work is supported by the International Cooperation Projects of the National Key R\&D Program (No. 2022YFE0127300), the
Science Foundation of Yunnan Province (No. 202401AS070046), the National Natural Science Foundation of
China (Nos. 11933008 and 11922306), the Agency of innovative development under the Ministry of higher education, science
and innovation of the Republic of Uzbekistan. Project (AL-5921122128): "Investigations and observations of special LAMOST
eclipsing binaries based on telescopes in Uzbekistan and China", CAS "Light of West China" Program and Yunnan Revitalization
Talent Support Program.The new CCD photometric data were obtained with the Maidanak Astronomical Observatory 1.5 m telescope.
This work has made use of data from the All-Sky Automated Survey for Supernovae (ASAS-SN), Catalina Real-Time Transient
Survey (CRTS), The Zwicky Transient Facility (ZTF), the Large-Sky-Area Multi-Object Fiber Spectroscopic
Telescope (LAMOST), the Transiting Exoplanet Survey Satellite (TESS), and Gaia and the authors thank the teams
for providing open data.

\bibliographystyle{mnras}
\bibliography{bib} 
\bsp	
\label{lastpage}
\end{document}